\documentclass[usenatbib]{mn2e}
\usepackage{amssymb}
\usepackage{epsfig}

\begin{document}

\title[{\sl Chandra} and {\sl XMM-Newton} Observations of
NGC\,4214]{{\sl Chandra} and {\sl XMM-Newton} Observations of
NGC\,4214: The Hot Interstellar Medium and the Luminosity Function of
Dwarf Starbursts}
 
\author[J.M. Hartwell et al.]
{Joanna M. Hartwell,$^1$ Ian R. Stevens,$^1$ David
K. Strickland,$^2$\thanks{{\it Chandra} Fellow}   
\newauthor Timothy M. Heckman$^2$ and Lesley K. Summers$^1$ \\
$^1$School of Physics and Astronomy, University of Birmingham, Edgbaston, Birmingham B15 2TT \\
E-mail: jmh@star.sr.bham.ac.uk, irs@star.sr.bham.ac.uk,
lks@star.sr.bham.ac.uk \\ 
$^2$ Department of Physics and Astronomy, The Johns Hopkins
University, 3400 North Charles Street, Baltimore, MD 21218, USA \\
dks@pha.jhu.edu, heckman@pha.jhu.edu}

%\author{Joanna Hartwell}

\maketitle

\begin{abstract}

We present results from {\sl Chandra} and {\sl XMM-Newton} X-ray
observations of NGC\,4214, a nearby dwarf starburst galaxy containing
several young regions of very active star-formation. Starburst regions
are known to be associated with diffuse X-ray emission, and in this case
the X-ray emission from the galaxy shows an interesting morphological
structure within the galaxy, clearly associated with the central regions
of active star-formation. Of the two main regions of star formation in
this galaxy, X-ray emission associated with the older is identified
whereas little is detected from the younger, providing an insight into the
evolutionary process of the formation of superbubbles around young
stellar clusters. The spectra of the diffuse emission from the galaxy can be
fitted with a two temperature component thermal model with $kT=0.14$\,keV
and 0.52\,keV, and analysis of this emission suggests that NGC\,4214
will suffer a blow-out in the future. 

The point source population of the galaxy has an X-ray luminosity
function with a slope of $-0.76$. This result, together with those for
other dwarf starburst galaxies (NGC\,4449 and NGC\,5253), was added to a
sample of luminosity functions for spiral and starburst galaxies. The
slope of the luminosity function of dwarf starbursts is seen to be
similar to that of their larger counterparts and clearly flatter than
those seen in spirals. Further comparisons between the luminosity
functions of starbursts and spiral galaxies are also made.
 
\end{abstract}


\section{Introduction}

Hierarchical models of structure formation in the Universe suggest that
dwarf galaxies were some of the first objects to be formed
\citep*{Navarro_95}, and hence were amongst the earliest sites of star
formation. If dwarf galaxies in the local Universe are analogues of
those formed at high redshift then studying the structure and evolution
of local dwarf galaxies will be useful in understanding star formation
at high redshifts.

When starburst galaxies undergo intense periods of star formation many
supernovae occur in a relatively small region of the galaxy. The ejecta
from these combine with the winds and ejected material from the massive
stars in OB associations and Super Star Clusters (SSCs) in the
surrounding area to form a superbubble around the cluster which expands,
sweeping up surrounding medium to produce large volumes of hot, shocked
gas with $T\sim 10^8$K. The simulations of \citet{Suchkov_94} show that
X-ray emission from starburst galaxies is most likely to be due to the
shocks formed from the interaction between the starburst outflow and the
surrounding medium, rather than from the material within the winds. The
bubbles of hot gas expand along the steepest density gradient --
usually the minor axis of the galaxy, and will eventually expand out of
the galaxy's gravitational potential well. Indeed, H$\alpha$ images of
starburst galaxies show filaments extending along the minor axis
(\citealt{Marlowe_95}, \citealt{Lehnert_96a}). The low mass and hence
small gravitational potential well of dwarf galaxies means that it
more likely than in other
galaxies for material to escape into the surrounding intergalactic
medium and form galactic superwinds. In some models these
starburst-driven winds result in enough interstellar medium (ISM) being
ejected from the galaxy via galactic superwinds to prevent any further
star formation from being able to occur
\citep{De_Young_94}. \citet{Lehnert_96b} find that superwinds are most
commonly formed in galaxies which have extreme IR properties, such as
high IR luminosities and warm far-IR flux. Dwarf starburst galaxies can
therefore play an important role in enriching their surroundings with
heavy elements produced during starburst episodes (\citealt{Ferrara_00},
\citealt{Dekel_86}).

NGC\,4214 is a nearby dwarf starburst galaxy with an almost face-on
orientation. Using the tip of the red giant branch method the distance
to NGC\,4214 has been estimated by \citet*{Maiz-Apellaniz_02} as
$D=2.94$Mpc, and this is the value used throughout this paper. We note
that a distance of 2.7Mpc was derived by \cite{Drozdovsky_02} using the
same method. A few $\times10^7$ years ago the galaxy underwent a burst
of star formation during which around $5\%$ of the stellar mass of the
galaxy was formed \citep{Huchra_83}. Colour magnitude diagrams for the field
stars outside the main star-forming regions show evidence for both old
and young populations, and are typical for dwarf galaxies. The diagrams
are dominated by the ``red tangle'' -- an area of the diagram that contains
RGB, AGB and blue-loop stars -- which implies that there is an
underlying population of old stars \citep{Drozdovsky_02}. The basic
properties of NGC\,4214 are given in Table~\ref{tab:4214}. 

NGC\,4214 has been studied at many wavelengths, each of which provide an
insight into its structure and composition. These observations appear
to be show ambiguous results when considering whether NGC\,4214 has
undergone a recent interaction with a nearby galaxy. $I$~band observations of
NGC\,4214 show a smooth, symmetric disk \citep{Fanelli_97}. The outer
contours of the $I$~band emission are regular, which suggest that there
is no evidence that the galaxy has undergone a recent
merger. NGC\,4214 is H{\small I}-rich like many other dwarf
starbursts. In contrast, observations of H{\small I}~emission by \citet{Allsopp_79} show
an extension towards a nearby irregular galaxy NGC\,4190 suggesting a possible
 tidal interaction between the two galaxies could have triggered
the starburst in NGC\,4214. Observations in the far ultra-violet (FUV) shows the
presence of knots of emission at the centre of the galaxy embedded in
diffuse emission \citep{Fanelli_97}. These knots denote the presence of
OB associations which have also been observed at optical wavelengths
({\sl e.g.} \citealt{Huchra_83}, \citealt{Leitherer_96},
\citealt{MacKenty_00}).
 
More recent observations of NGC\,4214 using the Hubble Space Telescope
({\sl HST}) have shown evidence for a two-stage starburst
\citep{MacKenty_00}, producing SSCs and OB associations. These
observations show that the galaxy has a complicated morphology and is
dominated by two large H{\small II} complexes often referred to as the
North West (NW) and South East (SE) complexes (these structures are also
referred to as NGC\,4214-I and NGC\,4214-II in the literature as
well). In addition many knots of H$\alpha$ emission are observed along
with structurally varying diffuse gas surrounding the main regions of
emission. Recent CO observations \citep{Walter_01} show three main
sites of molecular emission, two of which correspond to the NW and SE
complexes, with an additional site at which star formation has yet to
commence. 

In optical appearance NGC\,4214 is similar to other dwarf
irregular galaxies, however \citet{Maiz-Apellaniz_99} suggest that it
has an unusually thin galactic disk with a thickness of around 200pc;
outflows from the galaxy extending this by a further 200pc or
so. Flatter galaxies preferentially
undergo blow-out which is when supernovae explosions blow through the
disk of the galaxy, forming a way for gas from subsequent explosions to
escape without having much affect on the surrounding gas
(\citealt{De_Young_94}, \citealt{Mac_Low_99}). 

%An H{\small{I}} shell has been observed \citep{Allsopp_79} which may be due
%to gas escaping from the thin disk and forming structures away from the
%centre of the galaxy. 

\begin{table}
\begin{center}
\caption{Properties of NGC\,4214}
\begin{tabular}{ c c c } 

\hline \hline
Parameter & Value & Ref. \\ \hline

$\alpha$[2000] & $12^{h}15^{m}39.26^{s}$ & \\
$\delta$[2000] & $+36^{\circ}19{'}33.4{''}$ & \\
Distance & 2.94Mpc & 1\\
1 arcsec & 855pc & \\
$D_{25}$ & $8.91'\times5.62'$ & 2 \\
Position angle & 42.2$^{\circ}$ & 2 \\
Metallicity & $\sim0.25$ & 3 \\
m$_B$ & 10.20\,mag & 4 \\
S$_{60\mu m}$ & 17.9\,Jy & 5 \\
S$_{100\mu m}$ & 29.0\,Jy & 5 \\ 
$L_{FIR}$ & 9.1$\times10^{41}\rm{erg\,s}^{-1}$ & 6 \\
SFR & 0.04\,$M_{\odot}\,\rm{yr}^{-1}$ & 7 \\ \hline \hline
%SFR & 0.5--1.0$M_\odot\,\rm{yr}^{-1}$ & 6 \\ \hline \hline
\end{tabular}
\end{center}
References: (1) \citet{Maiz-Apellaniz_02}; (2) LEDA database
(available at http://leda.univ-lyon1.fr); (3) \citet{Kobulnicky_96};
(4) NED (NASA Extragalactic
Database, available at http://nedwww.ipac.caltech.edu); (5)
\citet{Soifer_89}; (6) see Section~\ref{sect:disc_ptsrcs},
Table~\ref{table:lumfunc} for more details; (7) using the $L_{FIR}$
relation from \citet{Kennicutt_98b}. 

\label{tab:4214}

\end{table}

%Another nearby dwarf starburst is NGC\,1569, a dwarf irregular galaxy
%which has recently undergone a wide-scale burst of star formation that
%involved most of the optical galaxy \citep{Hunter_00}. Due to its
%proximity (around 2.2Mpc) it is one of the best studied starburst
%galaxies. A study of the chemical enrichment of the interstellar medium
%by the massive stars in the starbursts of NGC\,4214 and NGC\,1569 has
%been carried out by \citet{Kobulnicky_96} and \citet{Kobulnicky_97}
%respectively. Using optical data they search for the inhomogeneities in
%the interstellar medium that are expected due to the presence of the
%winds and supernovae from the large star clusters. An overabundance of
%oxygen is observed surrounding the SE complex in NGC\,4214, the youngest
%starburst region, probably due to recent supernovae. Other than this no
%evidence for local enrichment in either of the two galaxies is found,
%even though the data should be more than capable of detecting such
%inhomogeneities. \citet{Kobulnicky_97} favour the explanation that the
%elements that must be produced by the stellar clusters are hidden from
%optical observations due to the fact that they are in a highly ionized
%hot phase. If this is indeed the case then X-ray observations of these
%galaxies should yield data that explain what is happening to these heavy
%elements.

The launch of the {\sl Chandra} Observatory has revolutionised X-ray
astronomy, providing an unprecedented spatial resolution of
$0.5''$. This resolution enables the identification of point sources
down to fainter luminosities than those seen with the previous X-ray
telescopes such as {\sl ROSAT} and {\sl Einstein}. It is now possible to
study the diffuse X-ray emission from galaxies without severe
contamination from unresolved point sources, allowing a more accurate
definition of its properties. The {\sl XMM-Newton} telescope launched in
1999 combines both reasonable spatial resolution with very high
sensitivity. This allows the temperature and the elemental abundances of
hot, diffuse gas to be studied in great detail. The combination of data
from both instruments promises a much deeper understanding of starburst
galaxies.

Prior to the launch of {\sl Chandra} and {\sl XMM-Newton} X-ray data
from several dwarf starburst galaxies had been obtained from {\sl ROSAT}
observations (examples of {\sl ROSAT} data on dwarf starburst galaxies
can be found in \citet{Stevens_98a} and \citet{Stevens_98b}). However,
{\sl ROSAT} lacked both the sensitivity and resolution to be able to
disentangle the point source emission from X-ray binaries from the
diffuse emission, which severely limits the interpretation of the data.

This analysis of NGC\,4214 adds to a small sample of published results
from {\sl Chandra} and {\sl XMM-Newton} observations of dwarf
starbursts. One significant example is the analysis of a {\sl Chandra}
observation of NGC\,1569
\citep*{Martin_02}. With the resolution of {\sl Chandra} it is possible
to study not just the global properties of the X-ray emission but also
the properties of different spatial components. \citet{Martin_02} found
a strong correlation between the X-ray and H$\alpha$ emission, and found
a high $\alpha$-element to Iron abundance ratio ($\alpha$/Fe), which is
consistent with a large amount of Type {\small II} supernova ejecta
contribution \citep{Martin_02}. It is concluded from this study that
elements can be released from dwarf galaxies into the surrounding
intergalactic medium. Further comparison is presented in
Section~\ref{sect:xray+ir}. In addition to NGC\,1569 published data is
available from {\sl Chandra} observations of the dwarf starbursts
NGC\,4449 \citep{Summers_03a}, NGC\,3077 \citep{Ott_03} and NGC\,5253
\citep{Summers_03b}.

In Section~\ref{sect:obs} we describe the {\sl Chandra} and {\sl
XMM-Newton} observations of NGC\,4214 and the data reduction methods
used. In Section~\ref{sect:ptsrcs} the properties of the point sources
are analysed, and in Section~\ref{sect:diff} the properties of the
diffuse X-ray emission are investigated. The morphology of this emission
is discussed and compared the H$\alpha$ emission from the central
regions of the galaxy. In Section~\ref{sect:discussion} the luminosity
function of NGC\,4214 is compared to those of spiral and other starburst
galaxies and the properties of the diffuse emission and their
implications for the formation of superbubbles are discussed. Finally in
Section~\ref{sect:concl} the main conclusions are summarised.

\begin{figure*}
\centering
\includegraphics[scale=1.0]{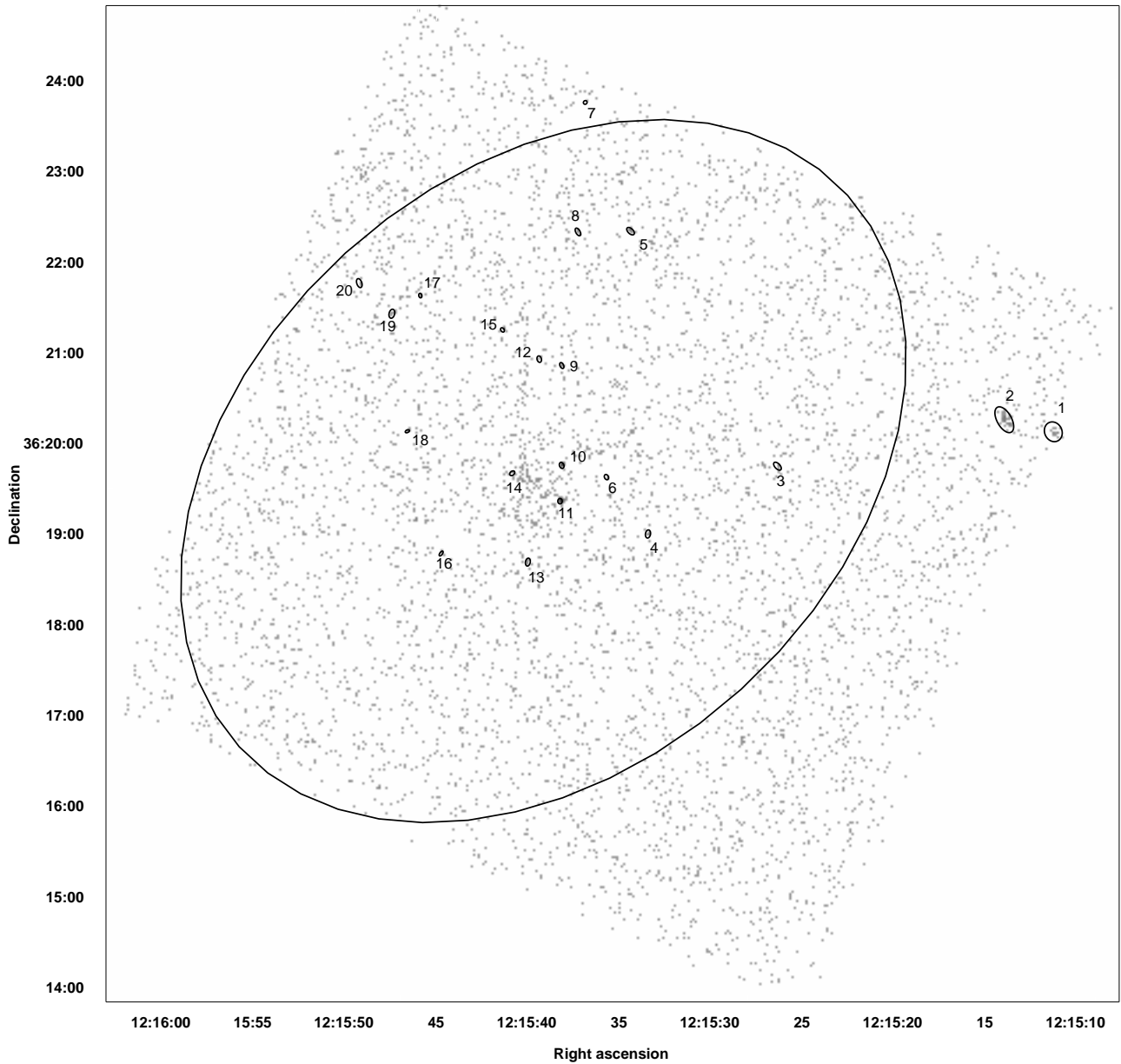}
\caption{The NGC\,4214 field, as seen with {\sl Chandra}, showing the
whole of the ACIS-S3 chip ($8.5'\times8.5'$), binned up by a factor of
2. NGC\,4214 can be seen as a region of diffuse emission at the centre
of the image. The large ellipse represents the $D_{25}$ ellipse of the
galaxy; the smaller ellipses show the location and extent of the
discrete sources as found by {\sl wavdetect}.}
\label{fig:whole_chip}
\end{figure*}

\section{Observations and Analysis} 
\label{sect:obs}

\subsection{Chandra Observations }
\label{sect:ch_anl}

NGC\,4214 was observed using the Advanced CCD Imaging Spectrometer
(ACIS) instrument on board the {\sl Chandra} satellite on 22nd October
2001, using the ACIS-S configuration.  The observation was a single
exposure of 26.4ksec, although cleaning the data left just over
16.5ksec. The entire galaxy is included on the back-illuminated ACIS-S3
chip hence only this chip was used in the analysis. The data reduction
was undertaken using {\sl Chandra} Interactive Analysis of Observations
(CIAO) version 2.2.1 software analysis tools\footnote{see {\sl Chandra}
data analysis methods, published by the {\sl Chandra} X-ray Center on
http://asc.harvard.edu/ciao/}. The level 2 event files were filtered to
reject all events with energies outside the range of 0.3--10.0\,keV
and to include event grades 0,2,3,4 and 6. The
background light curve was analysed, and all periods of the observation
containing flaring or periods of high background above the $3\sigma$
level were removed using the {\sl lc$\_$clean.sl} script provided by the
{\sl Chandra} X-ray Center (CXC).

The point sources were located using the {\sl wavdetect} programme, and
spectral analysis was carried out using {\sl XSPEC} (version 11.2.0). To
correct for the low energy quantum efficiency (QE) degradation caused by
molecular contamination of the ACIS optical blocking filters
\citep{Plucinsky_02} the CXC supplied {\sl corrarf} code was used to
correct the effective area file. This is dependent on the number of days
between the launch date of {\sl Chandra} and the observation, in this
case day 818.

Fig.~\ref{fig:whole_chip} shows the data from the ACIS-S3 chip of the
NGC\,4214 field. The $D_{25}$ ellipse of NGC\,4214 has semi-major and
semi-minor axes of 4.5$'$ and 2.8$'$ respectively, corresponding to an
area of $\sim39$ square arcminutes. The $D_{25}$ ellipse is shown in
Fig.~\ref{fig:whole_chip}, as are the 20 detected point sources. The
detection and analysis of the point sources are discussed in more detail
in Section~\ref{sect:ptsrcs}, while the diffuse emission associated with
the central regions of NGC\,4214 is discussed in
Section~\ref{sect:diff}.

\begin{figure}
\centering
\includegraphics[scale=0.47]{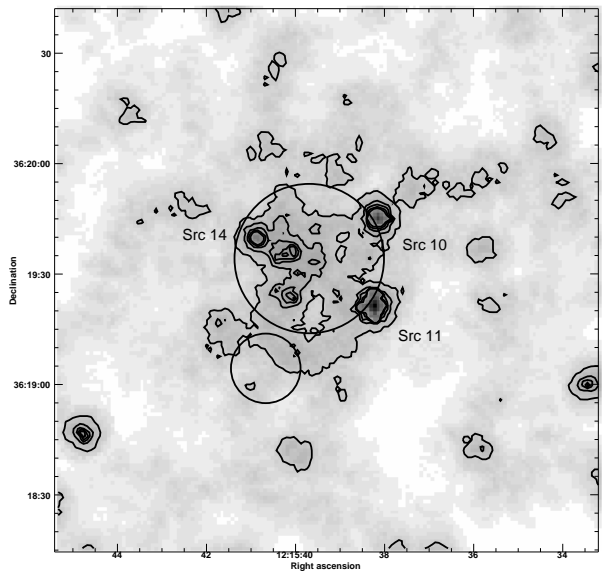}
\caption{The X-ray emission from NGC\,4214 observed with {\sl
Chandra}, binned by a factor of 2
and adaptively smoothed. The region shown is $2'\times2'$. The
X-ray emission can be seen to coincide with the H$\alpha$ emission
(see text), the NW and SE complexes of which are
depicted by the two circles. The three central point sources,
Srcs~10, 11 and 14, are clearly seen. See Section~\ref{sect:halpha} for
further discussion on the correlation between the X-ray and H$\alpha$
emission.}
\label{fig:gal+cont}
\end{figure}

Zooming into the central regions of NGC\,4214, Fig.~\ref{fig:gal+cont}
shows the {\sl Chandra} X-ray data for the central $2'\times 2'$. For
this image the data has been binned by a factor of 2 and then adaptively
smoothed. This central region contains 3 point sources (Srcs~10, 11 and
14), as well as diffuse emission.  The extent of the NW complex, as seen
in H$\alpha$ emission, is depicted by the large circle in
Fig.~\ref{fig:gal+cont}. The smaller blue circle represents the extent
of the SE complex of H$\alpha$ emission, which does not show as
much X-ray emission. The central X-ray emission, and correlations with
H$\alpha$ emission is discussed in detail in Section~\ref{sect:halpha}.

\subsection{XMM-Newton Observations}

The {\sl XMM-Newton} observation of NGC\,4214 was taken on November 22nd
-- 23rd 2001 for a total duration of 14ksec. The data was analysed using
the {\sl XMM-Newton} Science Analysis System (SAS) version 5.3.3
standard processing analysis. The data from the two MOS cameras and the
PN were processed using the the processing chains {\sl emchain} (MOS)
and {\sl epchain} (PN) to remove both bad pixels and bad events, and the
data were checked for periods containing background flaring. After these
periods were removed there was around 13ksec of data left for the two
MOS cameras, and 12ksec for the PN.

When the {\sl Chandra} and {\sl XMM-Newton} observations were compared
it was clear that there was a significant positional offset of around
15${''}$. To investigate the pointing accuracy of the observations
sources common to both observations were correlated to sources found in
the APM (Automatic Plate Measuring) catalogue\footnote{Available at
http://www.ast.cam.ac.uk/$\sim$mike/apmcat/} (including sources found on
chips other than the ACIS-S3 chip used in this paper). Two matches were
found, and the positions of the sources in the {\sl Chandra} observation
matched those in the APM very well, whereas both of the {\sl XMM-Newton}
sources displayed and offset of around $15{''}$. In addition an archived
{\sl ROSAT} observation shows increased X-ray emission corresponding
with the position of the X-ray emission observed by {\sl
Chandra}. Throughout this paper the positions obtained from the {\sl
Chandra} data are adopted, and the {\sl XMM-Newton} data has been
shifted accordingly.

\begin{figure}
\centering
\includegraphics[scale=0.47]{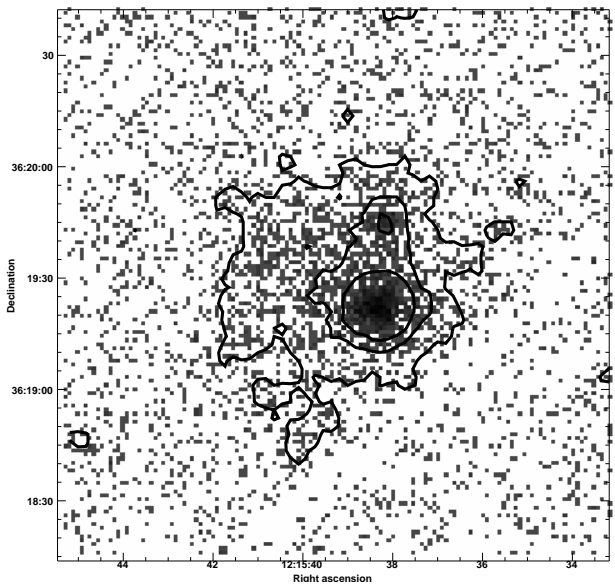}
\caption{The diffuse emission from NGC\,4214 as observed with {\sl
XMM-Newton}. The data is from a mosaiced image of the MOS1, MOS2 and PN
instruments. The region shown is $2'\times2'$, the same as that of the
{\sl Chandra} image shown
in Fig.~\ref{fig:gal+cont}. In the central regions of the galaxy at
least 2 of the 3 central point sources
seen with {\sl Chandra} are apparent, though blended with diffuse
emission. In addition, the diffuse emission can be seen to extend
towards the East.}
\label{fig:xmm_gal}
\end{figure}

In Fig.~\ref{fig:xmm_gal} we show a mosaiced image from the three {\sl
XMM-Newton} cameras of the central regions of the NGC\,4214 field. The
diffuse emission seen in the {\sl Chandra} data is also visible in this
observation, and the emission is clearly not symmetric, with an
extension to the North East. In the {\sl Chandra} observation 3 point
sources were identified in the central regions of NGC\,4214, while in
the {\sl XMM-Newton} data they are much less clearly resolved. Because
of the confusion between point sources and diffuse emission in the
central regions of NGC\,4214 we shall concentrate on the {\sl Chandra}
data for much of the paper.

\section{Point Source Analysis}
\label{sect:ptsrcs}

In this section we describe the detection and investigation of the point
sources in the NGC\,4214 field.  The point source detection tool {\sl
wavdetect} was used on the {\sl Chandra} ACIS-S3 chip, with a detection
threshold of $5.41\times10^{-7}$ (proportional to the inverse of the
number of pixels in the image) -- this level means that approximately
1 spurious point source detection is expected in the field. Using this method, a
total of 20 point sources were found (as shown in
Fig.~\ref{fig:whole_chip}) and the positions of these sources are listed
in Table~\ref{table:position}. Of these sources a total of 17 fall
within the $D_{25}$ ellipse of the galaxy which is depicted as the large
ellipse in Fig.~\ref{fig:whole_chip}. Those sources outside the $D_{25}$
ellipse are Srcs~1, 2 and 7. It is worth noting that the apparent
elongated shape of 1 and 2 are due to the shape of the instruments' PSF
towards the edge of the chip. 

The point source detection regions
generated by {\sl wavdetect} are output as ellipses which contain
99.7\% ($3\sigma$) of the source counts. For each detection a
background region was defined as an annulus centred on the source with
an inner radius just larger than the maximum axis of the detection
ellipse and an outer radius 3 pixels larger than this. Local
background regions are useful as they allow for any local diffuse
emission on which the point source emission is superimposed to be
taken into account. The
background subtracted counts of the sources are also listed in
Table~\ref{table:position}.

\begin{table*}
\caption {The positions of the 20 point sources detected in the {\sl
Chandra} observation of the NGC\,4214 field. In addition the number of
counts (background subtracted) detected for each source in the total
energy range and in the soft, medium and hard energy bands are
shown. Errors are calculated on the counts using Poissonian
statistics. For sources with $<60$ counts the unabsorbed flux has been
calculated by assuming an absorbed power law model with an exponent of
1.8 and an absorption equivalent to the Galactic value of
$N_H=1.5\times10^{20}\rm{cm}^{-2}$ (see text for full explanation). The
fluxes of the 4 sources with $>60$ counts have been obtained from
individual fits to the sources (see Section~\ref{sect:ind_ptsrcs}).}

\begin{center}
\begin{tabular}{ c c c c c c c c } \hline
 & RA (h m s) & Dec ($^o$ $'$ $''$) & net counts & Soft band & Medium band & Hard band & Flux \\
 Source  & J2000 & J2000 & ($\pm$ error) & ($\pm$ error) & ($\pm$ error) & ($\pm$ error) & erg~cm$^{-2}$s$^{-1}$ \\ 
 & & & 0.3--8.0\,keV & 0.3--0.8\,keV & 0.8--2.0\,keV & 2.0--8.0\,keV &  0.3--8.0\,keV \\ \hline

1$^a$ & 12:15:11.17 & 36:20:07.0 & 27.0$\pm$5.2 & 6.0$\pm$2.5 & 9.0$\pm$3.0 & 12.0$\pm$3.5 & $9.78\times10^{-15}$ \\ 
2$^{a,b}$ & 12:15:13.93 & 36:20:15.0 & 63.7$\pm$8.6 & $-1.9\pm$2.3 & 31.0$\pm$5.6 & 34.6$\pm$6.2 & $5.83\times10^{-14}$ \\
3 & 12:15:26.35 & 36:19:44.5 & 10.6$\pm$3.4 & 5.6$\pm$2.5 & 4.0$\pm$2.0 & 1.0$\pm$1.0 & $3.8\times10^{-15}$ \\
4 & 12:15:33.44 & 36:18:59.7 & 10.0$\pm$3.2 & 4.0$\pm$2.0 & 6.0$\pm$2.5 & -- & $3.6\times10^{-15}$ \\
5 & 12:15:34.38 & 36:22:20.0 & 35.6$\pm$6.0 & 3.0$\pm$1.7 & 18.6$\pm$4.4 & 14.0$\pm$3.7 & $1.3\times10^{-14}$ \\
6 & 12:15:35.71 & 36:19:37.3 & 6.7$\pm$2.7 & 6.7$\pm$2.7 & -- & -- & $2.4\times10^{-15}$\\
7$^a$ & 12:15:36.86 & 36:23:45.2 & 3.0$\pm$1.7 & -- & 2.0$\pm$1.4 & 1.0$\pm$1.0 & $1.1\times10^{-15}$ \\
8 & 12:15:37.26 & 36:22:19.6 & 15.6$\pm$4.0 & 5.0$\pm$2.2 & 11.0$\pm$3.3 & $-0.4\pm$0.4 & $5.7\times10^{-15}$ \\
9 & 12:15:38.15	& 36:20:51.2 & 19.7$\pm$4.5 & 1.0$\pm$1.0 & 15.7$\pm$4.0 & 3.0$\pm$1.7 & $7.1\times10^{-15}$ \\
10$^b$ & 12:15:38.16 & 36:19:45.1 & 70.0$\pm$8.5 & 11.3$\pm$3.5 &
45.7$\pm$6.8 & 13.0$\pm$3.6 & $3.41\times10^{-14}$  \\
11$^b$ & 12:15:38.25 & 36:19:21.4 & 547.8$\pm$23.5 & 94.4$\pm$9.8 & 308.7$\pm$17.7 & 144.7$\pm$12.1 & $2.97\times10^{-13}$ \\
12 & 12:15:39.39 & 36:20:55.4 & 8.6$\pm$3.0 & 5.6$\pm$2.5 & 3.0$\pm$1.7 & -- & $3.1\times10^{-15}$ \\
13 & 12:15:40.01 & 36:18:41.1 & 8.6$\pm$3.0 & 5.0$\pm$2.2 & 2.0$\pm$1.4 & 1.6$\pm$1.5 & $3.1\times10^{-15}$ \\
14 & 12:15:40.86 & 36:19:39.9 & 15.8$\pm$4.2 & 1.0$\pm$1.0 & 9.8$\pm$3.4 & 5.0$\pm$2.2 & $5.7\times10^{-15}$ \\
15$^b$ & 12:15:41.40 & 36:21:14.7 & 151.1$\pm$12.3 & 26.5$\pm$5.2 & 86.0$\pm$9.3 & 38.5$\pm$6.3 & $7.27\times10^{-14}$ \\
16 & 12:15:44.76 & 36:18:46.8 & 10.6$\pm$3.3 & 1.0$\pm$1.0 & 2.8$\pm$1.7 & 6.8$\pm$2.7 & $3.8\times10^{-15}$ \\
17 & 12:15:45.90 & 36:21:37.5 & 7.8$\pm$2.8 & -- & 5.8$\pm$2.5 & 2.0$\pm$1.4 & $2.8\times10^{-15}$ \\
18 & 12:15:46.61 & 36:20:07.8 & 10.0$\pm$3.2 & 10.0$\pm$3.2 & -- & -- & $3.6\times10^{-15}$ \\
19 & 12:15:47.46 & 36:21:25.4 & 11.0$\pm$3.3 & 4.0$\pm$2.0 & 4.0$\pm$2.0 & 3.0$\pm$1.7 & $4.0\times10^{-15}$ \\
20 & 12:15:49.24 & 36:21:45.7 & 11.6$\pm$3.5 & 1.0$\pm$1.0 & 1.6$\pm$1.5 & 9.0$\pm$3.0 & $4.2\times10^{-15}$ \\ \hline

\end{tabular}
\flushleft
$^a$  sources which fall outside the $D_{25}$ ellipse of the galaxy. \\
$^b$  flux obtained from fitted spectrum, see Section~\ref{sect:ind_ptsrcs}.
\end{center}
\label{table:position}
\end{table*}

For all of the detected point sources, the detected counts were
extracted for four different energy bands; firstly over the whole energy
range (0.3--8.0\,keV), then in a soft (S) band (0.3--0.8\,keV) a medium
(M) band (0.8--2.0\,keV) and finally a hard (H) band
(2.0--8.0\,keV). Background counts were taken from the background
annulus corresponding to each source for each band, and were subtracted
from the total. The background subtracted values for each band are shown
in Table~\ref{table:position}.  For two sources, namely Srcs~2 and 8,
there is a negative number of counts in one of the energy bands due to
the poor statistics.

Just 4 sources have $>60$ counts; Srcs~2, 10, 11 and 15. This is
sufficient for a spectral fit to be made, and these sources are
discussed further in Section~\ref{sect:ind_ptsrcs}. For the remaining 16
sources the Portable, Interactive Multi-Mission Simulator, {\sl
PIMMS}\footnote{Available at http://asc.harvard.edu/toolkit/pimms.jsp},
was used to predict a flux assuming the exponent of the power law
component $\Gamma$=1.8 and a column density,
$N_H$, equal to the Galactic value of $N_H=1.5\times10^{20}\rm{cm}^{-2}$.

We find that the luminosity lower-limit for a reliable detection in the
{\sl Chandra} observation, assuming that the source is at the distance
of NGC\,4214, is $L_X\sim 10^{36}$erg~s$^{-1}$. 

%(The conversion factor
%for flux to luminosity, assuming a distance of 2.94Mpc, is
%1.03$\times10^{51}$ for NGC\,4214).
 
Comparing the point sources in the NGC\,4214 observation with the
results from the Chandra Deep Field-South observation reported on by
\citet{Rosati_02}, we expect on average $\sim 5$ of the sources seen in
this observation to be background objects and $\sim 3$ of the sources
within the NGC\,4214 $D_{25}$ ellipse to be background objects. This
suggests that the majority of sources detected in this observation are
associated with NGC\,4214.

\subsection{Hardness Ratios}
\label{sect:hrs}

Most of the sources detected in this observation are very faint; there
are only two sources with $>150$ counts and just two more with
$>60$. However, for faint sources it is possible to derive some spectral
information using hardness ratios. Hardness ratios provide information
about both the absorption and the temperature of point sources. For
data of sufficient quality it is possible to use a hardness ratio plot
to distinguish between
different classes of sources such as supernova remnants, supersoft
sources, X-ray binaries and AGN;
\citet{Prestwich_03} describe the classification of sources using this
method.

\begin{figure}
\centering
\includegraphics[scale=0.4]{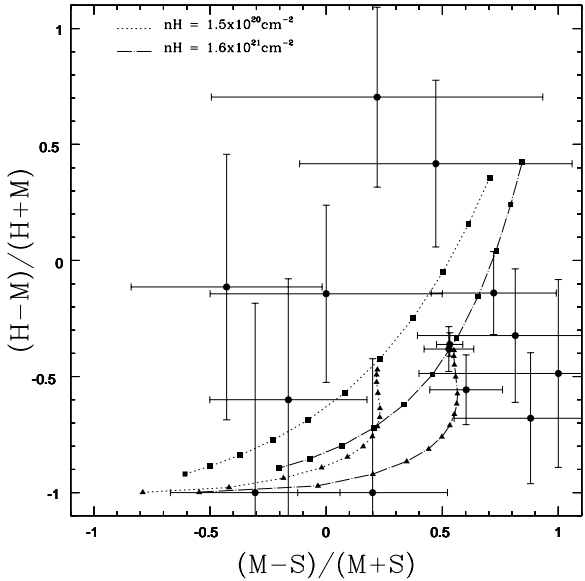}
\caption{The hardness ratios calculated for the point sources within the
$D_{25}$ ellipse, omitting 4 sources (see Section~\ref{sect:hrs} for
explanation). The energy bands used are: S$=$0.3--0.8\,keV,
M$=$0.8--2.0\,keV and H$=$2.0--8.0\,keV. The connected triangles
represent the colours predicted by an absorbed thermal bremsstrahlung
model with $kT$ varying from 0.2 (bottom left) to 2.0\,keV in increments
of 0.2\,keV, and from 2.0 to 3.6\,keV in increments of 0.4\,keV. The
connected squares represent the colours predicted from an absorbed power
law model with an exponent $\Gamma$ varying from 0.0 (top right) to 4.0
in increments of 0.4. Models have been calculated using the Galactic
absorption ($N_H=1.5\times10^{20}\rm{cm}^{-2}$), and also with the Galactic
absorption plus the absorbing column density for NGC\,4214
($N_H=1.6\times10^{21}\rm{cm}^{-2}$, see text for further explanation).}
\label{fig:hr}
\end{figure}

The counts in the S, M and H bands
were used to calculate 2 hardness ratios using
(H$-$M)/(H+M) and (M$-$S)/(M+S) \citep{Summers_03a}. Fig.~\ref{fig:hr}
shows the hardness ratio plot. Two sources had no counts in 2 of the
energy bands, and hence are not included in the plot; neither were the
two sources for which negative values were found in one of the energy
bands. 

{\sl PIMMS} was used to predict colours for point sources with both
thermal bremsstrahlung and power law models, with varying $kT$ and
$\Gamma$ respectively. In addition, models have been calculated with
both Galactic absorption, ($N_H=1.5\times10^{20}\rm{cm}^{-2}$), and
with the column density derived from a spectral fit of the X-ray
emission from the entire $D_{25}$ ellipse of NGC\,4214
($N_H=1.6\times10^{21}\rm{cm}^{-2}$). The details and further
parameters obtained from this fit are discussed in
Section~\ref{sect:sfr}. The theoretical tracks produced by these
results are also shown in Fig.~\ref{fig:hr}.

There is no real correlation seen between the two hardness ratios. There
appear to be two distinct groups of sources; those with the
smallest errors correspond to the sources with the largest number of
counts and fall in the same region of the diagram. These points all
fall to the right of the track obtained using the higher absorption
component implying that $N_H=1.6\times10^{21}\rm{cm}^{-2}$ is a lower
limit on the value of absorption for each source. None of the sources
within the $D_{25}$ ellipse are dominated by
counts in the hard band, and there are 4 sources that have no counts in
the hard band at all. This information agrees with the previous
assertions that the majority of the point sources observed are
associated with the galaxy and are not objects such as background
quasars.

\subsection{Individual Point Sources}
\label{sect:ind_ptsrcs}

Four sources (Srcs~2, 10, 11 and 15) have $>60$ counts, and these were
fitted individually using a power law model with an absorption component
frozen at the Galactic value of $N_H=1.5\times10^{20}\rm{cm}^{-2}$ and with
a further absorption component allowed to fit freely. The spectra were
grouped so that there were a minimum of 5 counts in each bin, with a
minimum of 10 for Src 11. Srcs~ 2 and 10 do not have sufficient counts
to allow the power law component to fit freely, and hence it was frozen
at a value of $\Gamma=1.8$. In the analysis of sources 11 and 15 this
parameter was allowed to vary freely. The results of the spectral fits
are shown in Table~\ref{table:ptsrcs}.

\begin{table*}
\caption {The components of the models fit to the 4 point sources with
$>60$ counts. Each was fitted with an absorbed power law model with
two absorption components; one frozen at the Galactic absorption value
of $N_H=1.5\times10^{20}\rm{cm}^{-2}$ and the other allowed to vary to
account for any additional local absorption. Column
2 shows the value of the second fitted absorption component and column
3 the exponent from the power law component. Column 4
gives the statistics for the fit, and column 5 the absorption
corrected luminosity for each source.} 
\begin{center}
\begin{tabular}{ c c c c c } \hline

Source & $N_H\,(\rm{cm}^{-2}$) & $\Gamma$ & $\chi_{\nu}^2$/d.o.f. & $L_X\,(\rm{erg}\,\rm{s}^{-1})$ \\ \hline

2 & $(8.0\pm^{11.2}_{8.0})\times10^{21}$ & 1.8(frz) & 0.57/10 & $(6.0\pm^{3.5}_{2.8})\times10^{37}$ \\
10 & $(1.8\pm^{1.8}_{1.3})\times10^{21}$ & 1.8(frz) & 1.12/10 & $(3.6\pm^{1.2}_{1.1})\times10^{37}$ \\
11 & $(1.8\pm^{0.6}_{0.7})\times10^{21}$ & 1.8$\pm^{0.3}_{0.2}$ & 0.8/44 & $(3.0\pm^{0.8}_{0.6})\times10^{38}$ \\ 
15 & $(1.0\pm^{1.2}_{0.2})\times10^{21}$ & 1.4$\pm^{0.5}_{0.3}$ & 1.2/24 & $(7.3\pm^{5.2}_{1.6})\times10^{37}$ \\ \hline

\end{tabular}
\end{center}
\label{table:ptsrcs}
\end{table*}
  
Cross-correlating with the APM catalogue 1 optical source was found to
correspond to an X-ray detection. It corresponds to Src~15, at a
separation of $0.22''$ in RA and $0.98''$ in Dec. The position of the
optical source was at $\alpha[2000]=12^{h}15^{m}41.38^{s}$,
$\delta[2000]=36^{\circ}21{'}15.7{''}$, compared to
$\alpha[2000]=12^{h}15^{m}41.40^{s}$,
$\delta[2000]=36^{\circ}21{'}14.7{''}$ for Src~15.

Src~10 corresponds to the position of a weak H$\alpha$ source observed
by \citet{MacKenty_00} (denoted as NGC\,4214-IIIs in that paper). A
source at this position was also observed by \citet{Fanelli_97} in an
$I$~band image of the galaxy, but not in the FUV. It is likely that this
source is at the dynamical centre of the galaxy, a view supported by the
H{\small I} observations of \citet{McIntyre_98} which place the
rotational centre of the galaxy close to the position of Src~10.

For a comparison, \citet*{Grimm_03} suggest that the number of high mass
X-ray binaries present in a galaxy is proportional to the star formation
rate, and use data from a number of galaxies to derive the following
relation between the number of bright X-ray sources (those with
$L_X>2\times10^{38}\rm{erg\,s}^{-1}$) and the SFR of the galaxy:

\begin{equation}
N(L>2\times10^{38}\rm{erg\,s}^{-1})=(2.9\pm0.23)\times
SFR\,[M_{\odot}\,\rm{yr}^{-1}] 
\end{equation}

Using the current SFR of $\sim 0.5-1.0 M_\odot \rm{yr}^{-1}$ calculated by
\citet{Thronson_88} it is predicted that in NGC\,4214 there should be
between 1 and 3 sources with $L_X>2\times10^{38}\rm{erg\,s}^{-1}$. We
detect one such source in this observation, Src 11. While many
ultraluminous X-ray sources (ULXs), which typically have X-ray
luminosities of $L_X\sim 10^{39}$ -- $10^{41}$erg~s$^{-1}$ have been
observed in other star-forming galaxies (e.g. the Antennae; see
\citealt{Zezas_02}), none are detected in NGC\,4214.

\subsection{The Luminosity Function}

Since the launch of {\sl Chandra} it has been possible to study the
X-ray point source populations in nearby galaxies in great
detail. Studying the luminosity function of a galaxy provides an insight
into its star formation history, and the luminosity functions of
different types of galaxy can be compared to look for common trends. The
point sources observed in NGC\,4214 are most likely X-ray binaries. An
X-ray binary remains active for a time which is dependent on the
lifetime of its companion star. In the case of a high mass X-ray binary
(HMXB) the companion is an OB star with a lifetime of around
$10^6-10^7$ years, however the low mass companions of low mass X-ray
binaries (LMXBs) usually have a longer lifetime, and so LMXBs remain
active for a longer period but at a lower luminosity than HMXBs
\citep{Wu_01}.  Luminosity functions for objects with a substantial
young population, such as
starburst galaxies, will be
dominated by the HMXBs with no significant contribution from LMXBs,
however as the HMXBs reach the end of their X-ray activity the LMXBs
will start to dominate \citep{Kilgard_02}. Any HMXBs observed must be
relatively young objects, and will have formed during the central
starburst of NGC\,4214. LMXBs have a lower luminosity limit of
$L_X\sim10^{36}$erg~s$^{-1}$ (\citealt{Schulz_99}, \citealt{Hertz_83}),
which is roughly the detection limit of this observation, but although some
of the fainter X-ray sources could be LMXBs, due to the relatively short period
since the starburst it is likely that the point source population
is dominated by HMXBs.

\begin{figure}
\centering
\includegraphics[scale=0.41]{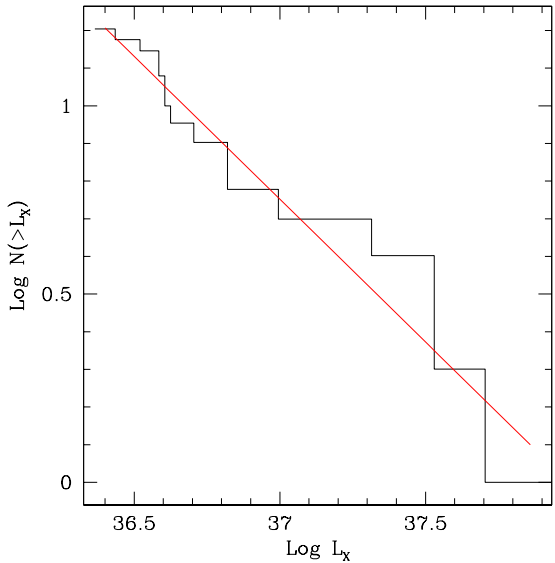}
\caption{The cumulative luminosity function for the 17 point sources within
the $D_{25}$ ellipse of NGC\,4214. The line depicts the power law
fitted to the data, with a slope of $-0.76\pm0.20$.}
\label{fig:logn}
\end{figure}

As discussed in Section~\ref{sect:ptsrcs}, the predicted fluxes for
the 17 point sources within the $D_{25}$ ellipse of the galaxy were
obtained by assuming a power law model with an exponent of $\Gamma =
1.8$, assuming that they are all at the distance of
NGC\,4214. Figure~\ref{fig:logn} shows the  $\log(N)-\log(L_X)$ plot for
these sources. The data are best fitted with a power law with a slope
of $-0.76\pm0.20$. It can be seen that below a luminosity of
$\log(L_X)\sim36.6$ the data appear to be flatter than the fitted
slope -- this is likely due to
an incompleteness in the detection of the point sources due to the
limits of the instrument. The luminosity function of NGC\,4214 in
comparison to those of other galaxies is
discussed further in Section~\ref{sect:disc_ptsrcs}.

\section{Diffuse Emission}
\label{sect:diff}

The faint diffuse emission from NGC\,4214 can be seen in the centre of
the {\sl Chandra} image shown in Fig.~\ref{fig:whole_chip}. It is clear
that the diffuse emission comes from a region far smaller than the
$D_{25}$ ellipse, as was shown in the blow-up of the central region in
Fig.~\ref{fig:gal+cont}.

The X-ray contours in Fig.~\ref{fig:gal+cont} show that there is a
bright ring of emission with a diameter of very roughly $0.6'$
(510pc). In addition, three point sources close to the edge of the ring
can be seen. The diffuse emission from the galaxy can be seen to a
greater extent in the {\sl XMM-Newton} image (Fig.~\ref{fig:xmm_gal});
in fact the radius of this emission is around $1.4'$, almost twice that
than detected by {\sl Chandra}. However the spatial resolution is not
good enough to clearly distinguish the point sources observed with {\sl
Chandra} from the rest of the central diffuse emission. Although the
detection limit of the {\sl Chandra} observation is similar to the
lower luminosity limit of LMXBs it is possible that if several sources
are present within a small area, for example in a cluster or
association, they will not be distinguishable as separate individual
sources and their emission is likely to contaminate that of the
observed diffuse emission.

\subsection{X-ray Analysis}
\label{sect:X-ray_anl}

The X-ray spectrum of the diffuse emission of NGC\,4214 from the {\sl
Chandra} observation was extracted using standard CIAO software, in an
energy range of 0.3--8.0\,keV, excluding the three point sources from
the analysis. The
background used was from the blank-sky background
files associated with the data set -- this allowed the background to be
taken from the same position in the instrument as the source
emission. The region used for extraction was an ellipse with axes of
$0.64'$ and $0.60'$ ($\sim 550\times 500$pc) chosen to encompass the
diffuse emission.

The diffuse emission spectrum contains 470 counts and was
fitted with several different models in an
attempt to obtain the best fit, each with a component, $wabs_G$, due
to Galactic absorption fixed at $N_H=1.5\times10^{20}\rm{cm}^{-2}$.

The model which provided the best fit to the data was
$wabs_G(wabs*mekal+wabs*mekal+po)$ which consists of two absorbed
thermal components (the $mekal$ components model the emission spectrum
from hot gas and include line emission from several elements) and a
power law component to account for any unresolved point sources. The
statistics of the fit give a reduced $\chi^2$ of 0.77 for 33 degrees of
freedom. The two thermal components both have absorptions of
$N_H=1.9\times10^{21}$cm$^{-2}$ and temperatures of 0.14\,keV and
0.52\,keV respectively, with the power law component having a photon
index of $\Gamma=1.56$. The unabsorbed flux of the diffuse
emission is $3.4\times10^{-13}$erg~s$^{-1}$cm$^{-2}$ (0.3--8.0\,keV), which
corresponds to an unabsorbed luminosity of
$L_X=3.5\pm_{3.0}^{32.5}\times10^{38}$erg~s$^{-1}$ (non-absorption corrected
$L_X=1.1\times10^{38}$erg~s$^{-1}$). Of this emission around $70\pm^{27}_{49}$\% was
due to the softer component, $16\pm^{41}_{11}$\% to the harder and $14\pm^{9}_{10}$\% to the power
law component. The errors on these percentage contributions show that
while there is a degree of uncertainty as to the percentage
contribution of the softer and harder thermal components, the division of the
emission between the combination of the two thermal components and the
power law component appears to be fairly well constrained.

The parameters of the fit, together with the errors at
the $90\%$ confidence level (1.64$\sigma$) are shown in
Table~\ref{tab:fit_param}. It is noted that if the abundances are
allowed to vary (as opposed to being frozen at solar) unphysical
values from the fit are obtained - this is most likely due to there
being too many free parameters for values to be constrained.

The spectral response of the ACIS CCDs is known to vary over time due
to both the evolution of the electronic gain of the CCDs, and the
variance of their charge transfer inefficiency (CTI). A correction for
the time dependence of the gain of the ACIS chips using the 
$corr\_tgain$ tool \footnote{contributed CXC software, see
http://asc.harvard.edu/cont-soft/software/corr\_tgain.1.0.html}, was
applied to the data. When the data were reanalysed with this
correction it was seen that the values of the absorption and
temperature components changed very little, although the slope of the
power law component of the spectrum fitted to the diffuse emission became
flatter. The effect of this correction was small, and $corr\_tgain$ is
not yet an official CXC product, so as such the results presented here
and used for further analysis are those from the uncorrected data.

\begin{figure}
\centering
\includegraphics[scale=0.35,angle=270]{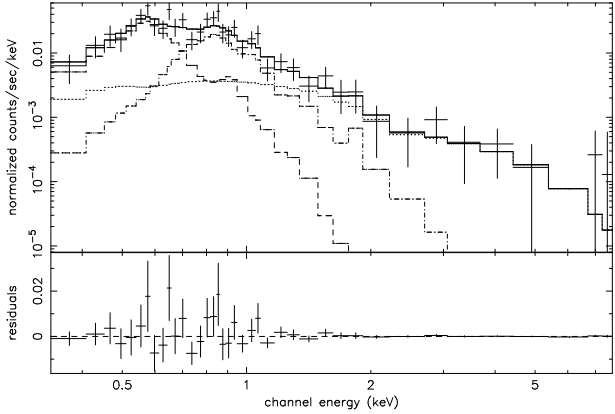}
\caption{The spectrum of the diffuse {\sl Chandra} X-ray emission from
NGC\,4214 (0.3--8.0\,keV), fit with a 2 temperature {\it mekal} model, giving
temperatures of 0.14\,keV and 0.52\,keV, and a power law component
with a photon index of $\Gamma=1.56$. The contributions of the
individual components are depicted by the dashed lines.} 
\label{fig:ch_diff_spectrum}
\end{figure}

\begin{figure}
\centering
\includegraphics[scale=0.35,angle=270]{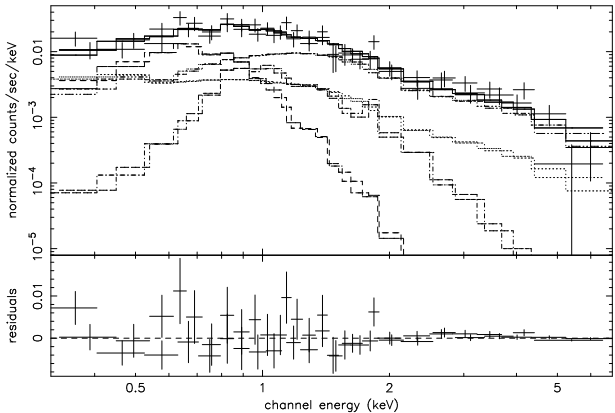}
\caption{The spectrum of the diffuse {\sl XMM-Newton} X-ray emission from
NGC\,4214 (from the two MOS cameras only). The spectra was fitted over
an energy range of 0.3--8.0keV with a 2 temperature {\it mekal} model with
to power law components to take into account unresolved point sources,
one fixed with the properties obtained from the {\sl Chandra} fits to
Srcs 10, 11 and 14 (see text for further explanation). The softer
thermal component is fit with a temperature of
0.23\,keV, the harder with 0.59\,keV.}
\label{fig:xmm_spectrum}
\end{figure}

Concerning the {\sl XMM-Newton} data, although {\sl XMM-Newton} is more
sensitive than {\sl Chandra} it is hampered by the fact that its spatial
resolution is poorer. From the {\sl Chandra} analysis it is known that
there are three point sources (Srcs 10, 11 and 14) amongst the diffuse
emission in the centre
of the galaxy, and these could be excluded from the {\sl Chandra}
analysis. However, in the {\sl XMM-Newton} observation these cannot be
distinguished from the diffuse emission with sufficient accuracy to
allow them to be excluded without removing significant parts of the
diffuse emission as well. To take into account these 3 sources an
additional absorbed power law component
with values frozen at $N_H=1.8\times10^{21}$cm$^{-2}$ and $\Gamma=1.8$
was included in the models (the values used were based on those
obtained from the fits to Srcs 10 and 11 -- see
Table~\ref{table:ptsrcs}). The spectra were again fitted with a
$wabs_G(wabs*mekal+wabs*mekal)$ model, with the additional component to take into
account the central point sources included. It was found that the spectra from
the two MOS cameras showed good agreement with the results obtained from
{\sl Chandra}; however, when the results from the PN were included a fit
with unphysical values was obtained. The discrepancy between the
different cameras is most probably due to the differences in calibration
between them. The fits were thus carried out using the MOS cameras
only. 

The fit to the {\sl XMM-Newton} data gives two absorption
components of $N_H=1.7\times10^{21}$cm$^{-2}$ and
$N_H=8.5\times10^{21}$cm$^{-2}$, a temperature for the softer
component of $kT=0.23$\,keV and for the harder of $kT=0.59$\,keV and a
power law component with a photon index of
$\Gamma=1.93$. The results of the fit, together with the errors on the
fitted values are given in Table~\ref{tab:fit_param}. The unabsorbed
luminosity was found to be
$L_X=6.0\pm_{1.6}^{1.4}\times10^{38}$erg~s$^{-1}$, higher than that obtained from
the {\sl Chandra} fit which is to be expected due to the inclusion of
point source emission. Indeed, when the calculated contribution of Srcs
10, 11 and 14 is subtracted from this value a luminosity of
$L_X\sim2.5\times10^{38}$erg~s$^{-1}$ is obtained which is in better
agreement with that from the {\sl Chandra} observation. The contribution
of the various components to this emission was found to be $19\pm^2_5\%$ for
the cold component, $34\pm^7_{11}\%$ for the hot, $9\pm^3_3\%$ due to the power law
component, and $38\pm^3_3\%$ due to the {\sl Chandra} sources 10, 11 an
14. The results from the {\sl XMM-Newton} data appear to broadly
confirm those obtained from {\sl Chandra}. 

\begin{table}
\caption{The parameters from the best fit model to both the {\sl Chandra}
and {\sl XMM-Newton} data. Each data set was fitted with a {\it
wabs$_G$*(wabs*mekal+wabs*mekal+po)} model; the {\sl XMM-Newton} data had an
additional power law component with an exponent of $\Gamma=1.8$ and
$N_H=1.8\times10^{21}$cm$^{-2}$ to take into account the unresolved
central point sources (Srcs 10, 11 and 14) -- see text. In each case {\it wabs$_G$}
was assumed to be the Galactic absorption value of
$N_H=1.5\times10^{20}\rm{cm}^{-2}$. The errors shown are those calculated
at the $90\%$ confidence level.}

\begin{tabular}{ c c c } \hline

 & {\sl Chandra Data} & {\sl XMM-Newton Data} \\ \hline
$N_{H1}$ ($10^{22}\rm{cm}^{-2}$) & 0.19$\pm_{0.19}^{0.53}$ & 0.17$\pm_{0.03}^{0.06}$ \\
$kT_1$ (keV) & 0.14$\pm_{0.04}^{0.06}$ & 0.23$\pm_{0.03}^{0.04}$ \\
$N_{H2}$ ($10^{22}\rm{cm}^{-2}$) & 0.19$\pm_{0.19}^{0.48}$ & 0.85$\pm_{0.13}^{0.27}$ \\
$kT_2$ (keV) & 0.52$\pm_{0.44}^{0.19}$ & 0.59$\pm_{0.19}^{0.19}$ \\
$\Gamma$ & 1.56$\pm_{1.11}^{0.76}$ & 1.93$\pm_{0.45}^{0.51}$ \\
$\chi_{\nu}^2$/d.o.f. & 0.77/31 & 1.11/40 \\
$L_X$ ($10^{38}\,\rm{erg\,s}^{-1}$) & $3.53\pm_{3.0}^{32.5}$ & $6.04\pm_{1.64}^{1.37}$ \\ \hline

\end{tabular}
\label{tab:fit_param} 
\end{table}

By making some basic assumptions it is possible to calculate
additional parameters of the gas components of NGC\,4214. The
following calculations were carried out using the results obtained
from the {\sl Chandra} analysis. \citet{Maiz-Apellaniz_99} suggest
that NGC\,4214, which we are observing face on, has an unusually thin
disk with a total height of around 200pc extended by outflows by perhaps
a further 200pc. As such a cylindrical geometry for the galaxy is
assumed with a thickness of 200pc and a diameter of $\sim0.6'$
(510pc) which corresponds to that of the observed emitting region. 

The mass of the emitting gas can be calculated from the normalisations
of the model components. The normalisation, $k$ is given by

\begin{equation}
k=\frac{10^{-14}}{4\pi D^2}\int n_{\rm{e}} n_{\rm_{H}} dV
\label{eqn:k}
\end{equation}

\noindent where $D=2.94$Mpc, $V$ is the volume of the emitting region,
and $n_{\rm{e}}$ and $n_{\rm_{H}}$ are the number densities of electrons
and hydrogen ions respectively. We assume that the gas is fully ionised
($n_{\rm{e}}\sim\,n_{\rm_{H}}$). The total mass of the emitting gas is
$\sim2.1\times10^5\eta^{1/2}$M$_{\odot}$ with the two thermal
components and one power law component contributing
$1.5\times10^5\eta^{1/2}$M$_{\odot}$,
$4.2\times10^4\eta^{1/2}$M$_{\odot}$ and
$1.8\times10^4\eta^{1/2}$M$_{\odot}$ respectively. We note the
dependence on $\eta$, the volume filling factor of the X-ray emitting
material -- see \citet{Strickland_00a} for a discussion of its
properties. We make no assumptions about th value of $\eta$, but
\citet{Strickland_00a} find that it could be quite low ($\sim0.1$) for
soft emission; if this is the case both the mass and energy content of
the gas are reduced.

Further parameters of the emitting gas can be calculated by making some
basic assumptions about its properties. Using the emission integral,
$EI$, from Equation~\ref{eqn:k}, (($k\times4\pi D^2)/10^{-14}$), the
electron density can be calculated using $n_e\sim(EI/V\eta)^{1/2}$,
assuming the emitting volume, $V$, calculated previously. The
dynamical pressure, $P$, of the
gas is obtained using $P\sim2n_ekT\eta^{-1/2}$, and the thermal energy, $E_{Th}$,
of the gas from $E_{Th}\sim3n_ekTV\eta^{1/2}$. Finally the cooling time,
$t_{cool}$, of the gas is calculated using $t_{cool}\sim3kT\eta^{1/2}/\Lambda
n_e$ where $\Lambda = L_X/EI$. The results of these calculations for the
two components of the gas are shown in Table~\ref{tab:two_comp_par}.

\begin{table}
\caption{Parameters calculated from the best fit properties from the
{\sl Chandra} analysis of the diffuse emission. For assumptions used
in the calculations see text.}

\begin{tabular}{ c c c } \hline

Emission & Softer & Harder \\ 
 & component & component \\ \hline
$kT$ (keV) & 0.14 & 0.52 \\
$T$ (K) & 1.62$\times10^6$ & 6.04$\times10^6$ \\
$L_X$ ($\rm{erg\,s}^{-1}$) & 2.48$\times10^{38}$ & 5.57$\times10^{37}$ \\
$n_e$ ($\rm{cm}^{-3})\times(1/\sqrt\eta$) & 0.10 & 0.03 \\
$E_{Th}$ (erg) $\times(\sqrt\eta$ & 1.44$\times10^{53}$ & 1.61$\times10^{53}$ \\
$M$ ($M_{\odot})\times(\sqrt\eta$) & 1.82$\times10^5$ & 5.50$\times10^4$ \\
$P$ (dyn $\rm{cm}^{-2})\times(1/\sqrt\eta$) & $4.61\times10^{-11}$ & $5.16\times10^{-11}$ \\
$t_{cool}$ (yr)$\times(\sqrt\eta$) & 1.86$\times10^7$ & 9.32$\times10^7$ \\ \hline

\end{tabular}
\label{tab:two_comp_par}

\end{table}

\begin{figure*}
\includegraphics[scale=0.8]{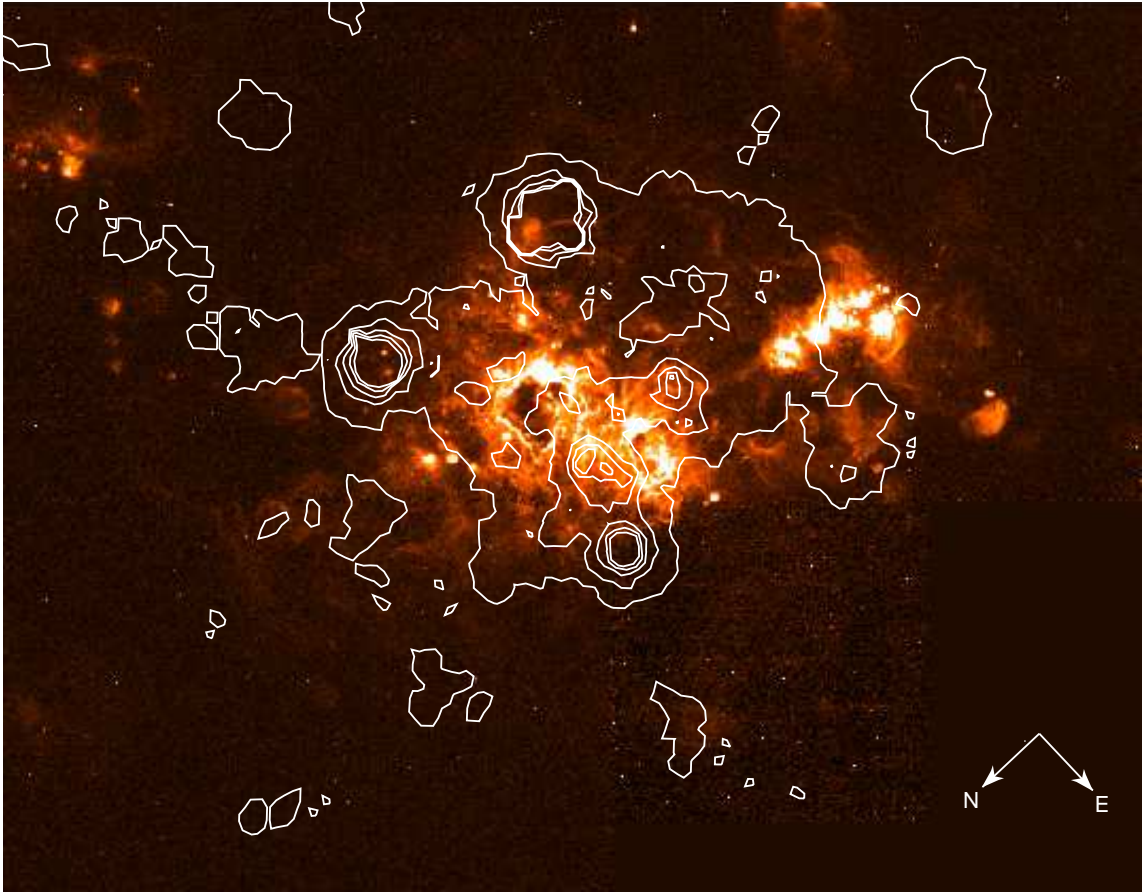}
\caption{The {\sl Chandra} observation of the central regions of
NGC\,4214. X-ray contours from the binned, smoothed
image are superimposed onto an H$\alpha$ image. In places regions of
enhanced X-ray emission clearly correspond to regions of
increased H$\alpha$ emission, but that in others there appears to be
no correlation. This H$\alpha$ image was kindly provided by Jes\'{u}s
Ma\'{\i}z-Apell\'{a}niz from data published in MacKenty et
al. (2000) and also available from http://www.stsci.edu/$\sim$jmaiz.}
\label{fig:halpha+cont}
\end{figure*}

\subsection{Comparison with Optical Observations}
\label{sect:halpha}

As has been noted before, excluding the point sources, the bulk of the
X-ray emission from NGC\,4214 comes from the central regions of the
galaxy. The centroid of the X-ray emission (located at
$\alpha[2000]=12^{h}15^{m}39.77^{s}$,
$\delta[2000]=36^{\circ}19{'}28.9{''}$, see Fig.~\ref{fig:gal+cont})
coincides with the NW star-formation complex, marked out by
H$\alpha$ emission, observed by \citet{MacKenty_00}.

Fig.~\ref{fig:halpha+cont} shows the H$\alpha$ emission from NGC\,4214
observed by \citet{MacKenty_00} with the X-ray contours from the {\sl
Chandra} data overlaid. This image shows clearly that there is more
X-ray emission associated with the NW star-forming complex than
with the SE complex. This is perhaps to be expected as the NW
complex is the more evolved and more massive of the two, hence it
contains more supernovae, and any superbubble will have had more time to
develop. Although the bulk of the X-ray emission is centred around the
point sources in the NW complex there is emission extending away in a
southerly direction which gives the impression of having a ring-like
shape. This emission is most likely solely due to the ongoing star
formation in the NW complex, with the shocked gas from stellar winds and
supernovae preferentially expanding into regions of the ISM with lower
ambient density. The geometry of the emission is somewhat reminiscent to
that of a so-called `champagne phase' (see \citealt{Yorke_86} for a
review).  There is little X-ray emission corresponding with the
position of the SE complex, which is further conformation that this is
the younger of the two star formation regions. It appears that
substantial superbubbles have not yet had time to form, and hence the
region is likely not yet able to inject significant energy into the
surrounding medium.
 
Src~10 corresponds to the position of a star cluster believed to be the
dynamical centre of the galaxy (as discussed in more detail in
Section~\ref{sect:ind_ptsrcs}). A young SSC, of age
$\sim3.0-3.5\,\rm{Myr}$, has been identified in
optical images at $\alpha[2000]=12^{h}15^{m}39.44^{s}$,
$\delta[2000]=36^{\circ}19{'}35.0{''}$ (\citealt{Leitherer_96},
\citealt{MacKenty_00}), and it appears to reside in an H$\alpha$
cavity. At this position there is increased
emission in the X-ray suggesting the presence of cluster type emission.

\section{Discussion}
\label{sect:discussion}

\subsection{Point Sources}
\label{sect:disc_ptsrcs}

The luminosity functions of galaxies can now be studied in great
detail due to the fact that {\sl Chandra} is able to detect many
individual X-ray point sources within the galaxies. The slope of the
luminosity function can be used as an indicator of some of the general
properties of the host galaxy. A recent study of 32 nearby spiral and
elliptical galaxies by \citet{Colbert_03} shows a clear distinction
between the slopes of the luminosity functions of these two types of
galaxies, with those of the spiral galaxies being much steeper than
those of the ellipticals.

In those galaxies where a large number of point sources has been
observed the luminosity functions are often fitted with power laws, for
example NGC\,4449 \citep{Summers_03a}, NGC\,5253 \citep{Summers_03b},
M31 \citep{Supper_97}, and M81 \citep{Tennant_01}.

If a continuous uniform star formation process occurs then the
luminosity function of the sources should appear as unbroken, but when
a starburst is triggered new HMXBs will be formed
which will appear as a break in the function. It is therefore not
unlikely that a break in the luminosity function will be present in
all starburst galaxies, the position of which will decrease in
luminosity with time, perhaps providing an indicator to the time of
previous bursts in the galaxy. It is possible that no break is seen in
NGC\,4214 due to the limited statistics and the small number of point
sources observed; if there is contamination of the luminosity function
by LMXBs these could mask a break in the HMXB population when there is
such a small sample being considered.

The results for the X-ray luminosity function presented here for
NGC\,4214 can be combined with those of other dwarf starbursts and then
further compared with results from a wider sample of galaxies for a range
of parameters (for example \citealt{Kilgard_02}, see below).  The
relevant parameters (and sources) for each galaxy used in this study are
shown in Table~\ref{table:lumfunc}.

The blue luminosity $L_B$ was calculated using \citet{Tully_88} 

\begin{equation}
\log L_B(L_\odot) = 12.192-0.4B_T+2\log D
\label{eqn:opt}
\end{equation}
with $D$ in Mpc, and $L_{FIR}$ calculated
using the 60$\mu$m and 100$\mu$m fluxes \citep{Devereux_89}
\begin{equation}
L_{FIR}(L_\odot) = 3.65 \times 10^5[2.58S_{60\mu m} + S_{100\mu m}]D^2
\label{eqn:FIR}
\end{equation}

\citet{Kilgard_02} examined the point source luminosity functions of 7
galaxies; 4 spirals and 3 starbursts. The 3 `normal' starbursts, M82,
NGC\,253 and the Antennae (NGC\,4038/9), have flatter luminosity
functions than the spirals, with slopes of $-$0.50, $-$0.81 and $-$0.53
respectively compared to a range of $-1.07$ to $-1.30$ for the spirals
(see Table~\ref{table:lumfunc}). We supplement this sample with results
for the dwarf starbursts galaxies NGC\,4449 \citep{Summers_03a},
NGC\,5253 \citep{Summers_03b} and NGC\,4214 (this paper).

\begin{figure}
\centering
\includegraphics[scale=0.4]{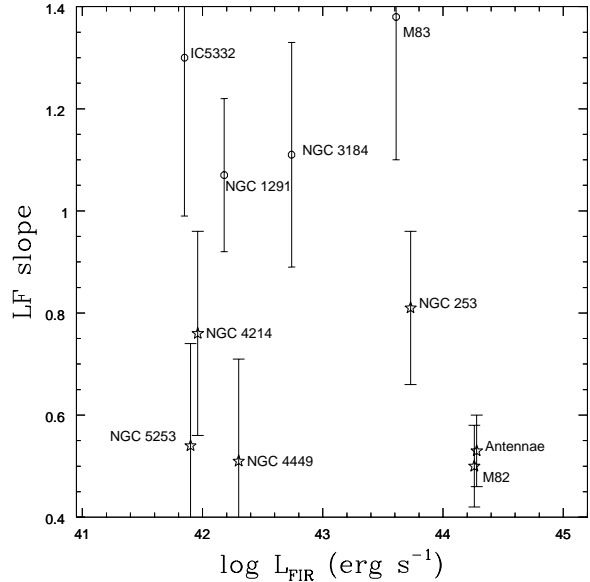}
\caption{The luminosity function slope plotted against $L_{FIR}$
for the \citet{Kilgard_02} sample of 4 spirals and 3 starbursts, with
the results for 3 dwarf starbursts (NGC\,4214, NGC\,4449 and
NGC\,5253) added. The spiral galaxies are denoted as circles in the
plot; the starbursts as stars.}
\label{fig:lf}
\end{figure}

The far infrared luminosity $L_{FIR}$ of a galaxy can be used as an
indication of a young stellar population and of its current rate of star
formation \citep{Kennicutt_98a}. Thus, if the luminosity function slope
is dependent on the rate of star formation, then we might expect a
correlation between the luminosity function slope and $L_{FIR}$.
Fig.~\ref{fig:lf} shows the luminosity function slope plotted versus
$L_{FIR}$, and it is obvious that the starburst galaxies have a flatter
slope than the 4 normal spiral galaxies, suggesting that the higher the
star formation rate the flatter the luminosity distribution.  This is
probably due to the fact that there is a larger proportion of young
bright sources ({\sl e.g.} HMXBs) in star-bursting regions than is
present in spiral galaxies \citep{Wu_01}.  However, the dwarf starbursts
also have lower luminosity function slopes, but because the dwarf
starburst galaxies have $L_{FIR}$ values comparable to the spirals there
is no clear segregation on the basis of $L_{FIR}$.

\begin{figure}
\centering
\includegraphics[scale=0.4]{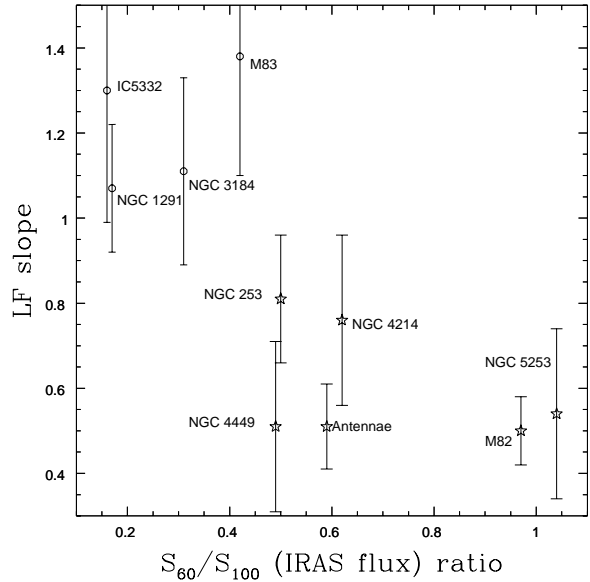}
\caption{The luminosity function slope plotted against the ratio of
the 60$\mu$m to 100$\mu$m flux for 4 spirals (circles) and 6 starbursts
(stars). A division can be seen between the two types of galaxies,
both in the luminosity function slope and in the flux ratio.}
\label{fig:f60_f100}
\end{figure}

A clearer segregation can be seen in Fig.~\ref{fig:f60_f100}, which
shows the X-ray luminosity function slope plotted against the ratio of
the 60$\mu$m ($S_{60 \mu m}$) to 100$\mu$m ($S_{100 \mu m}$) flux for
the spiral and starburst galaxies. \citet{Lehnert_96b} state that if
$S_{60 \mu m}/S_{100 \mu m} \gtrsim 0.5$ then it is more likely that a
galactic superwind will occur. As is clear from Fig.~\ref{fig:f60_f100},
this criterion is satisfied for all the starburst galaxies in the
sample. A superwind is indeed observed in  both M82
(e.g. \citealt{Strickland_97}, \citealt{Lehnert_99}) and NGC\,253
(e.g.\citealt{Strickland_00b}). In NGC\,5253 multiple superbubbles
are observed in the X-ray \citep{Strickland_99} and H$\alpha$
filaments are seen extending away from the galaxy
\citep{Marlowe_95}. From NGC\,4449 both X-ray \citep{Summers_03a} and
H$\alpha$ \citep{Marlowe_95} emission from a superbubble is observed,
but a large H{\small I} halo surrounding the galaxy may prevent the
escape of material into the surrounding intergalactic medium. Using
the {\sl Chandra} X-ray data it has been calculated that NGC\,4214 is likely to
undergo a blow-out; this is discussed in more detail in
Section~\ref{sect:xray+ir}. As the Antennae are two merging galaxies
as yet no superwind has formed due to the disruption from the ongoing
interaction. 

\begin{figure}
\centering
\includegraphics[scale=0.4]{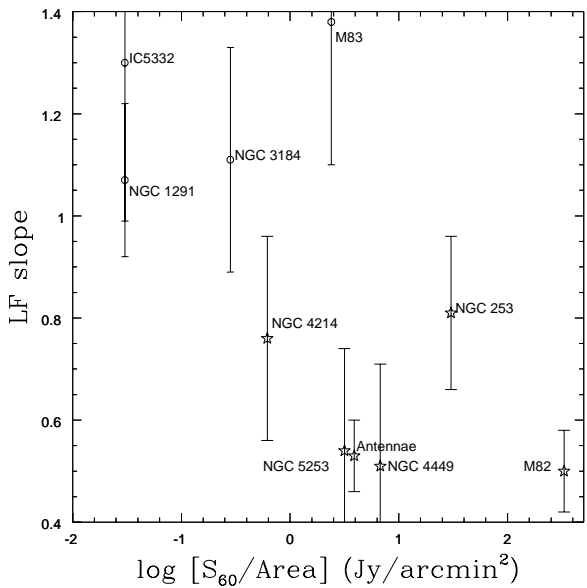}
\caption{The luminosity function slope plotted against the 60$\mu$m
flux divided by the area of the $D_{25}$ ellipse. (Spiral galaxies are
denoted as circles; starbursts as stars).} 
\label{fig:d25}
\end{figure}

Fig.~\ref{fig:d25} shows the luminosity function slope against the
60$\mu$m flux scaled to the area of the $D_{25}$ ellipse as a measure of
the star formation rate per unit area. Unsurprisingly, this tends to be
higher for the starburst galaxies than for the spirals (with the
exception of M83), and again we see a segregation in luminosity function
slope with star-forming activity. M83 has been classified as a nuclear
starburst galaxy, and {\sl HST} observations show a large number of
young, massive clusters in the nuclear region
\citep{Harris_01}. \citet{Kilgard_02} state that most of the detected
point sources used in the analysis of the luminosity function of M83 are
from regions outside the central star-bursting region. The fact that
only the disk population is analysed means that it can be grouped with
the other spirals in the sample when considering the luminosity function
slope. However, other values such as the optical luminosity ($L_B$) and
the far infrared luminosity ($L_{FIR}$) will have this central region
included which will affect the results. Indeed, inspection of the
60$\mu$m IRAS image of M83 shows a large variation of emission across
the galaxy, with the highest contribution coming from the centre. For
analysis purposes, however, the flux from the entire galaxy was
used. This explains why M83 seems to
agree with the properties of spirals in respect to the luminosity
function, but for other properties relating to the whole galaxy it is
more similar to what is seen in the starbursts. The luminosity
function of M82, another starburst galaxy with localised star
formation, contains point sources from the entire galaxy 
including the regions of star formation. Unlike M83 the
luminosity function slope of M82 shows agreement with those of the
other starburst galaxies. 

\begin{figure}
\centering
\includegraphics[scale=0.4]{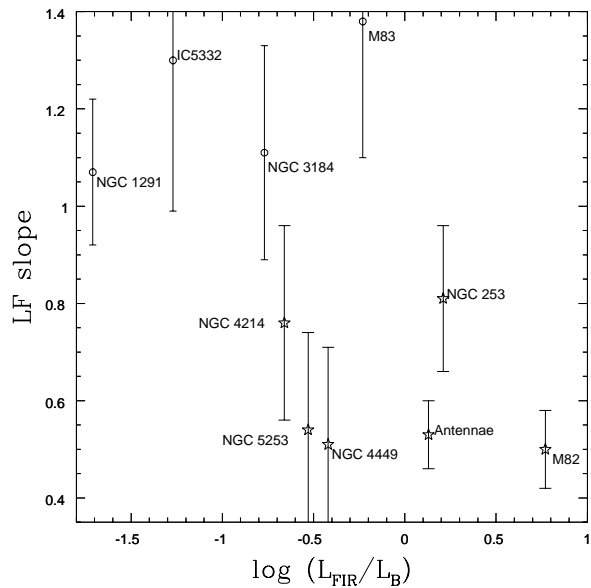}
\caption{The luminosity function slope for 4 spirals (circles) and 6
starbursts (stars)
plotted against $\log (L_{FIR}/L_B)$. This shows that in this sample
although in general the starbursts have a higher $L_{FIR}/L_B$ ratio
there is no clear division between the two types of galaxies.}
\label{fig:lfirlb}
\end{figure}

Typically $L_{FIR}$ scales with $L_B$ for normal galaxies, but in strong
starbursts $L_{FIR}$ is seen to increase more rapidly than
$L_B$. Fig.~\ref{fig:lfirlb} shows the slope of the luminosity function
plotted against $\log [L_{FIR}/L_B]$. In general, the starburst
galaxies in the sample have a higher $[L_{FIR}/L_B]$ ratio, which is to
be expected, but if we exclude M83 again there is still evidence of a
change in luminosity function slopes at a value of $\log L_{FIR}/L_B\sim
-0.75$.  M83 has the highest $[L_{FIR}/L_B]$ ratio of the spiral sample;
this could again be due to the fact that it has had some recent star
formation.

We note in passing that we can compare the total number of sources in
each galaxy with $\log L_X \geq 37.5$ (a luminosity level which provides
a compromise in that sources this faint are detected in more distant and
massive galaxies such as the Antennae and sources this bright are
detected in smaller and nearby dwarfs such as NGC\,4214 and
NGC\,5253). If we plot this number against $L_B$ then we find a linear
correlation with on average $0.2-1$ X-ray
sources with $L_X>10^{37.5}$~erg~s$^{-1}$ for each $10^8 L_\odot$ in
$L_B$.

\begin{table*}
\caption {Luminosity function data for a sample of nearby
galaxies; 3 dwarf starbursts, 4 starbursts and 3 spirals. Much of the
data is from the sample of \citet{Kilgard_02},
along with the data from NGC\,4214, NGC\,4449 and NGC\,5253 mentioned in the
text. $L_B$ and $L_{FIR}$ are estimated using Eqn.~\ref{eqn:opt} and
\ref{eqn:FIR}. The galaxy area is estimated from ($\pi\,a\,b$)cos$\theta$ where $a,\,b$
are the galaxy semi-major and
semi-minor axes of the $D_{25}$ ellipse, and $\theta$ is the
inclination of the galaxy, obtained from the LEDA database.}
\begin{center}
\begin{tabular}{ c c c c c c c c c} \hline
 & Distance & LF slope & 60$\mu$m flux & 100$\mu$m flux & $S_{60 \mu
m}/S_{100 \mu m}$ & $\log {L_{FIR}}$ & $\log {L_B}$ & $D_{25}$ area \\
Galaxy & (Mpc) & & (Jy) & (Jy) & & (erg s$^{-1}$) & (erg s$^{-1}$) & (arcmin$^2$) \\ \hline
NGC\,4214 & 2.9 & 0.70 & 17.9$^a$ & 29.0$^a$ & 0.62 & 41.96 & 42.62 & 29.2\\
NGC\,4449 & 2.9 & 0.51 & 36.0$^b$ & 73.0$^b$ & 0.49 & 42.30 & 42.72 & 5.3 \\ 
NGC\,5253 & 3.2 & 0.54 & 10.5$^c$ & 29.4$^c$ & 1.04 &  41.90 & 42.43 & 3.3\\
 & & & & & & & & \\
M82 & 5.2 & 0.50 & 1313.5$^a$ & 1355.4$^a$ & 0.97 & 44.26 & 43.49 & 4.0 \\
NGC\,253 & 3.0 & 0.81& 931.7$^a$ & 1861.7$^d$ & 0.50 & 43.73 & 43.52 & 31.1 \\
Antennae & 25.5 & 0.51 & 48.7$^a$ & 82.0$^a$ & 0.59 & 44.28 & 44.15 & 12.6 \\ 
 & & & & & & & & \\
NGC\,1291 & 8.6 & 1.07 & 1.8$^d$ & 10.1$^d$ & 0.17 & 42.18 & 43.89 & 70.3 \\
M83 & 4.7 & 1.38 & 266.0$^d$ & 638.6$^d$ & 0.42 & 43.61 & 43.84 & 110.8 \\
NGC\,3184 & 8.7 & 1.11 & 8.9$^a$ & 29.0$^a$ & 0.31 & 42.74 & 43.51 & 32.0 \\
IC\,5332 & 8.4 & 1.30 & 0.8$^c$ & 5.1$^c$ & 0.16 & 41.85 & 43.12 & 25.3 \\ 
\hline

\end{tabular}
\flushleft
$^a$ \citet{Soifer_89} \\
$^b$ \citet{Thronson_87} \\ 
$^c$ \citet{Moshir_90} \\ 
$^d$ \citet{Rice_88} \\

\end{center}
\label{table:lumfunc}
\end{table*}

It is worth noting that NGC\,4214 was analysed with a correction for the
low energy quantum efficiency degradation of {\sl Chandra} (see
Section~\ref{sect:ch_anl}), which gave slightly different values than
when it was analysed without this correction. The rest of the data was
obtained from previously published material which was analysed before
the correction was available and it is not known how much this will have
affected the validity of the results. However, this problem is getting
worse with time and will therefore have had less of an effect on
observations such as these which were taken early in {\sl Chandra}'s
lifetime, and so it is assumed that it is still valid to compare the
results obtained.

\subsection{Star Formation Rate of NGC\,4214}
\label{sect:sfr}

\citet*{Ranalli_03} propose that the X-ray luminosities of star-forming
galaxies can be used as an indication of the SFR of the galaxy. They
derive the following relation between the 2-10\,keV X-ray luminosity and
the SFR: 

\begin{equation}
SFR (M_{\odot}\,\rm{yr}^{-1}) = 2.0\times10^{-40}L_{2-10\,\rm{keV}}
\end{equation} 

The spectrum of the total X-ray emission within the $D_{25}$ ellipse of
the galaxy was extracted from the {\sl Chandra} data and fitted with an
absorbed thermal component
plus power law model within {\sl XSPEC}. The absorbing column density
was divided into 2 components, with one frozen at the Galactic column
density $N_H=1.5\times10^{20} \rm{cm}^{-2}$ and the other allowed to
fit. The exponent of the power law component was frozen at a value of
$\Gamma = 1.8$ to take into account the point sources emission.  The
fitted column for the additional absorbing component, local to
NGC\,4214, was found to be $N_H=1.4\times10^{21} \rm{cm}^{-2}$, giving a
total column of $1.6\times10^{21} \rm{cm}^{-2}$.  This basic spectral
analysis provided an unabsorbed 2--10\,keV X-ray luminosity of $L_X =
3.4\times10^{38} \rm{erg\,s}^{-1}$. When combined with the relation from
\citet{Ranalli_03} this gives a value of 0.07$M_{\odot}\,\rm{yr}^{-1}$
for the SFR of NGC\,4214.

\citet{Kennicutt_98b} derives the following relation between SFR and
the FIR luminosity in a galaxy:

\begin{equation}
SFR (M_{\odot}\,\rm{yr}^{-1}) = \frac{L_{FIR}}{2.2\times10^{43}}
\end{equation} 

\noindent
Using a value of $L_{FIR}=9.1\times10^{41}\rm{erg\,s^{-1}}$ (estimated
using Eqn.~\ref{eqn:FIR}) the Kennicutt relation gives a SFR of
0.04$M_{\odot}\,\rm{yr}^{-1}$, in good agreement with the value
obtained using the X-ray properties of the galaxy.

\subsection{Diffuse X-ray Emission and IR Properties}
\label{sect:xray+ir}

In this observation we have been able to resolve point sources and
diffuse emission in the central regions of NGC\,4214. This in turn
enables us to reliably estimate, for the first time, the energy present
in the ISM and ultimately whether we expect an outflow from NGC\,4214.

Metals and gas are potentially ejected from dwarf starburst galaxies
into the surrounding intergalactic medium, powered by the strong stellar
winds from OB stars, and then after $\sim 3.5$Myr predominantly by
the energy injected by supernovae, which adds to the mass loss from the
galaxy. Indeed, observations of  NGC\,1569 \citep{Martin_02} show that
the diffuse X-ray emitting gas has a supersolar $\alpha$/Fe
ratio. These results suggest that the hot
diffuse gas has been enriched by the ejecta of explosions Type II
supernovae injecting heavy elements into the surrounding medium.
The winds from the stars and supernovae in clusters and stellar associations
will combine, sweeping up the surrounding interstellar medium, creating
a superbubble of hot, X-ray emitting gas. Surrounding the superbubble
will be a cooler shell of material which emits H$\alpha$ line
emission. If instabilities within the bubble occur they can cause the
shell to rupture, allowing gas to escape in the form of a superwind.

\citet{Weaver_77} model bubbles for single stars, and these results are
often used to model the formation of bubbles around clusters and associations
that contain many massive stars. However, this model must be used with
caution. In the case of NGC\,4214 there are several separate
superbubbles that have been observed, which correspond to the different
clusters and associations at the centre of the
galaxy. \citet{MacKenty_00} observe bubbles around the star forming
regions in the NW complex that are around 30\% smaller than expected
using the \citet{Weaver_77} model, and conclude that the model cannot be
applied in this case. \citet*{Summers_01} find that the predicted
expansion velocity for the dwarf starburst galaxy Markarian 33 from
using the Weaver model is less than that observed.

In their observations of NGC\,4214 \citet{Maiz-Apellaniz_99} observe no
bubbles surrounding the three star clusters in the SE complex. They
suggest that observations of the SE complex do not agree with a blow-out
having occurred, and that despite the Weaver model predicting that there
should be a superbubble with R$_{\rm{B}}\sim$70pc and
$v_{\rm{B}}\sim$14\,km~s$^{-1}$ it is likely that a superbubble does not
actually exist. \citet{MacKenty_00} suggest that there is a delay in the
formation of a superbubble in clusters of about 2~Myr due to the time it
takes for the high number of individual stellar bubbles produced to
combine and break out of the surrounding medium. If this is indeed the
case it would explain why X-ray emission is seen to predominantly
coincide with the position of the somewhat older NW complex, where
cluster superbubbles have had time to form, but not with the SE complex
where there is a lack of shocked, hot gas. The time delay in the
formation of these clusters could have an impact on estimations of the
amount of energy released by the galaxy \citep{Maiz-Apellaniz_99}. The
mechanical energy could be dissipated via the heating of interstellar
gas, which would lead to an overestimation of the energy released, as
would the fact that stellar synthesis models assume that the massive
stars would appear earlier than is in fact the case.

\citet{Mac_Low_88} derive the following criteria for whether a
blow-out from a galaxy is to be expected:

\begin{equation}
\Lambda_{mm} = 10^4 L_{mech,41}H^{-2}_{kpc}P^{-3/2}_4n^{1/2}_0
\end{equation}

\noindent where $L_{mech,41}$ is the mechanical energy luminosity in
units of $10^{41}$erg~s$^{-1}$, $H_{kpc}$ is the scale height of the
galaxy in kpc, $P_4$ is the initial pressure of the ISM
in units of $P/k = 10^4\,\rm{K\,cm}^{-3}$ and $n_0$ is the initial
density of the ISM. $\Lambda_{mm}$ is a measure of the rate of kinetic
energy injection, and if $\Lambda_{mm} > 100$ then a blow-out will
occur. From Table~\ref{tab:two_comp_par}, the total thermal content of
the X-ray emitting gas is around
3.1$\times10^{53 }$erg~s$^{-1}$. Assuming that the observed diffuse X-ray
emission comes from the most recent burst of star-formation associated
with the NW complex, with an age of $\sim3.5$Myr \citep{MacKenty_00}, we
estimate that $L_{mech}\sim 2.8\times10^{39}$erg~s$^{-1}$. The initial
pressure of the ISM is assumed to be equal to that of the Milky Way,
i.e. $P_4\sim1$. For plausible values of the initial gas density $n_0$
(such as the initial density being $10~\times$ greater than the current
derived densities in Table~\ref{tab:two_comp_par}, cf
\citealt{Summers_03a}) for $\Lambda_{mm} > 100$ and a blow-out to occur
then we require $H_{kpc}\lesssim 3$kpc. Obviously the derived upper limit
for the scale height necessary for a blow-out to occur is very weakly
dependent on the assumed value of $n_0$. The assumed pressure is
comparable to that of \citet{Mac_Low_99} who assume
$P_4\sim4\times10^3\rm{K\,cm}^{-3}$. Indeed, if the lower value of
\citet{Mac_Low_99} is used it makes little difference to the result
($H_{kpc}\lesssim 2$kpc). The estimated value of
$H_{kpc}$ is significantly larger than the optically observed dimensions
of the galactic disk \citep{Maiz-Apellaniz_99}, and as such we conclude
that in this galaxy a blow-out from the disk ISM is likely to occur.

We note that \citet{Wilcots_01} calculate $L_{mech}$ to be
$\sim10^{39}\rm{erg\,s}^{-1}$, a factor of ten smaller than the value
obtained from our analysis. However, if $L_{mech}$ is indeed a factor
of ten smaller this will reduce the scale height by a factor of
$\sqrt10$, which will still be larger than the observed size of the
optical disk.

\citet{Lehnert_96b} have studied a large sample of starburst galaxies,
looking at which properties make them conducive to galactic superwinds
being formed, and find that those galaxies with extreme IR properties
are far more likely to have a superwind present. One of the criteria
is that the galaxy has a ratio of $S_{60\mu m}/S_{100\mu m}\gtrsim
0.5$; NGC\,4214 has a value of 0.62, implying that a superwind is a
possibility. However, NGC\,4214 fails to meet the other two criteria;
that $L_{FIR}\gtrsim 10^{44} \rm{erg\,s}^{-1}$ and
$L_{FIR}/L_B\gtrsim2$. These results appear to suggest that there is not a high chance of
this galaxy forming and supporting a galactic superwind, however it
should be noted that the galaxies used by \citet{Lehnert_96b} to
derive these criteria were mainly spiral galaxies, and as such the
results may not be quite so relevant for a dwarf starburst such as NGC\,4214. 

In the sample of 9 galaxies used in this to study the properties of
luminosity functions of starbursts, all of the starbursts have
$S_{60\mu m}/S_{100\mu m}\gtrsim0.5$ (note that NGC\,4449 has a value
of 0.49). However, with the exception of M82, with 
$L_{FIR}/L_B = 5.9$, none meet the other criteria for the formation of
a superwind to be likely.

\section{Summary and Conclusions}
\label{sect:concl}

We have presented the results from both {\sl Chandra} and {\sl
XMM-Newton} observations of NGC\,4214, which show the diffuse X-ray
emission from this dwarf starburst galaxy. The high spatial resolution
of {\sl Chandra} has allowed the detailed structure of the X-ray
emission of NGC\,4214 to be observed for the first time. 17 point
sources were observed within the $D_{25}$ ellipse of the galaxy, and
further analysis of these suggest that the majority are HMXBs associated
with the galaxy as opposed to being background AGN.

The luminosity function of the point sources is similar to those seen in
other starbursts such as NGC\,5253 and NGC\,4449. Indeed, when combined
with those of NGC\,4449 and NGC\,5253 and added to the sample of
\citet{Kilgard_02} this data strengthens the argument that the
luminosity functions of starburst galaxies are flatter than those seen
in normal spirals, and indeed extends this result to include dwarf
starburst galaxies. It is seen that the starburst galaxies tend to have
a higher 60$\mu$m to 100$\mu$m flux ratio than spirals, a criteria for
being more likely to produce a galactic superwind
\citep{Lehnert_96b}. However, there is no clear segregation between the
spirals and the starbursts when the luminosity function slopes are
compared to the ratio of $L_{FIR}/L_B$. \citet{Lehnert_96b} state that
for a superwind to form a large excess of IR luminosities is needed,
and although this criteria is not met in our sample overall the
starbursts show a larger $L_{FIR}/L_B$ ratio than the spirals, as would
be expected. In general the starbursts have a larger amount of star
formation per unit area (using the $D_{25}$ ellipse area as a scaling
parameter).

The diffuse emission from the galaxy is observed in both the {\sl
Chandra} and {\sl XMM-Newton} observation, and is seen to almost twice
the extension in {\sl XMM-Newton} (radius $\sim1.4'$) than in {\sl
Chandra}. The data suggest the presence of gas at two separate
temperatures, with {\sl Chandra} analysis providing values for these of
around 0.3 and 4.6\,keV and a combined absorption corrected flux of
$3.4\times10^{-13}$erg~s$^{-1}$cm$^{-2}$ corresponding to an unabsorbed
luminosity of $L_X=3.5\times10^{38}$erg~s$^{-1}$. Using derived X-ray
properties the SFR of NGC\,4214 is calculated as
0.07$M_{\odot}\,\rm{yr}^{-1}$, in reasonable agreement with the value of
0.04$M_{\odot}\,\rm{yr}^{-1}$ derived using $L_{FIR}$.

The X-ray emission coincides with regions of H$\alpha$ emission and
hence is coincident with star forming regions. The presence of X-ray
emission coincident with the NW complex, but not the SE
complex of H$\alpha$ emission agrees with previous speculation that
there is a time delay before the wind from a cluster of stars can break
out and start interacting with the surrounding medium. Using some basic
assumptions about the properties of the X-ray emitting gas is has been
calculated that it is likely that a blow-out will occur from the disk
of NGC\,4214, enriching the surrounding medium with materials from
this starburst.

\section*{Acknowledgements}

JMH and LKS acknowledge funding from PPARC studentships, and IRS from
a PPARC Advanced Fellowship. DKS is supported by NASA through {\sl
Chandra} Postdoctoral Fellowship Award Number PF0-10012, issued by the
{\sl Chandra} X-ray Observatory Center, which is operated by the
Smithsonian Astrophysical Observatory for and on behalf of NASA under
contract NAS8-39073. We thank Jes\'{u}s Ma\'{\i}z-Apell\'{a}niz for
supplying the H$\alpha$ image.

\bibliographystyle{mn2e}
\bibliography{jo}

\end{document}